\documentclass[
    pra,
    aps,
    reprint,
    amsmath,amssymb,
    a4paper,
    superscriptaddress,
    10pt
]{revtex4-2}

\usepackage{graphicx}
\usepackage{dcolumn}
\usepackage{bm}
\usepackage{ulem}
\usepackage{textcomp}
\usepackage{physics}
\usepackage[dvipsnames,table,xcdraw]{xcolor}
\definecolor{rmpblue}{HTML}{2e3092}
\usepackage[
    colorlinks=true,
    citecolor=rmpblue,
    linkcolor=rmpblue,
    filecolor=magenta,
    urlcolor=rmpblue
]{hyperref}
\usepackage{gensymb}
\begin{document}
\title[]{Metaphotonics for High-Harmonic Generation}

\author{Pavel Tonkaev}
 \email{pavel.tonkaev@anu.edu.au}
\affiliation{Research School of Physics, Australian National University, Canberra ACT 2601, Australia}

\author{Anton Rudenko}
\affiliation{
Laboratoire Hubert Curien, UMR CNRS 5516, Universit\'{e} Jean Monnet, 42000 Saint Etienne, France
}

\author{Ziwen Wang}
\affiliation{
Department of Physics and State Key Laboratory of Optical Quantum Materials, The University of Hong Kong, Hong Kong SAR, China
}

\author{Tran Trung Luu}
\affiliation{
Department of Physics and State Key Laboratory of Optical Quantum Materials, The University of Hong Kong, Hong Kong SAR, China
}

\author{Yuri Kivshar}
 \email{yuri.kivshar@anu.edu.au}
\affiliation{
Research School of Physics, Australian National University, Canberra ACT 2601, Australia
}

\begin{abstract}
We summarise the recent advances on the generation of high-order harmonics in optical metaphotonic structures such as isolated resonators and metasurfaces. In such subwavelength-patterned structures, extreme nonlinear effects are expected from the interaction between ultrafast laser pulses and structured planar surfaces which support various resonances. High-harmonic generation (HHG) is considered as an effective tool for realising extreme ultraviolet light sources and attosecond pulses, and it was observed previously in gases, liquids, and solids. Resonant metaphotonics can offer a novel sub-wavelength platform for the HHG effects, and it can provide new strategies for the design of efficient integrated light sources. We start our discussion from a brief overview of HHG in gases, liquids, and unstructured solids, and then move to summarising the recent experimental observations of HHG in individual resonant nanoparticles and resonant dielectric metasurfaces. We focus on different types of resonances (plasmonic vs. Mie resonances vs. bound states in the continuum), and also present the observation of non-integer power dependencies of the generated harmonics driven by strong resonances. We also mention current unsolved problems and identify new promising research directions involving HHG in metasurfaces.
\end{abstract}

\maketitle

\section{Introduction}

High-harmonic generation (HHG) is one of the manifestations of extreme nonlinear optics enabling the conversion of intense laser radiation into coherent light at multiples of the driving frequency, extending the cutoff photon energy from the ultraviolet to the extreme-ultraviolet and soft X-ray spectral regions~\cite{burnett1977harmonic,ferray1988multiple,paul2003quasi}. Originally discovered in atomic gases, HHG has become a cornerstone of attosecond science, providing compact coherent short-wavelength light sources and enabling direct access to ultrafast electron dynamics on femtosecond and attosecond timescales~\cite{paul2001observation,stolow2004femtosecond,cavalieri2007attosecond,corkum2007attosecond}. During the last decade, the observation of HHG in condensed matter systems has opened new opportunities for probing band-structure dynamics, strong-field transport, and many-body interactions in solids, while simultaneously offering a pathway toward compact solid-state coherent light sources~\cite{ghimire2011observation,schubert2014sub,luu2015extreme}.

Despite these remarkable advances, efficient HHG generally requires extremely high optical intensities because the nonlinear conversion efficiency decreases rapidly with increasing harmonic order. Achieving such intensities traditionally relies on amplified femtosecond laser systems delivering high pulse energies, which limits the scalability and integration of HHG technologies. Consequently, considerable effort has been devoted to engineering environments capable of enhancing the local optical field and strengthening light-matter interactions without increasing the incident laser power.

Metaphotonics has emerged as a powerful platform for addressing this challenge~\cite{baev2015metaphotonics,tonkaev2022all,zhang2024quantum,li2024ultrafast}. By structuring optical materials at the sub-wavelength scale, resonant nanophotonic systems, including isolated nanoresonators~\cite{rybin2024metaphotonics} and dielectric metasurfaces~\cite{vabishchevich2023nonlinear}, can confine electromagnetic fields into deeply sub-wavelength volumes while supporting optical resonances with exceptionally high quality factors and avoiding bulk propagation phase matching constraints. In particular, surface plasmon resonances \cite{han2016high, vampa2017plasmon}, Mie resonances~\cite{babicheva2024mie}, Fano resonances~\cite{vabishchevich2018enhanced}, guided-mode resonances~\cite{valencia2024enhanced}, and bound states in the continuum (BICs)~\cite{wang2025nonlinear} provide dramatic enhancement of local electromagnetic fields, enabling nonlinear optical processes that are otherwise inaccessible in bulk materials. Unlike plasmonic structures, all-dielectric metaphotonic platforms combine low optical losses with high damage thresholds and intrinsic optical nonlinearities, making them especially attractive for strong-field nonlinear optics~\cite{li2017nonlinear,zubyuk2021resonant,grinblat2021nonlinear,lin2025nonlinear}.

The application of metaphotonic concepts to HHG has rapidly evolved from enhancing harmonic conversion efficiencies to fundamentally controlling the underlying nonlinear interaction. Recent studies have demonstrated harmonic generation up to the eleventh order from resonant dielectric metasurfaces,\cite{liu2018enhanced,zograf2022high} giant enhancements of harmonic yield through quasi-bound states in the continuum~\cite{zalogina2023high,tonkaev2025unconventional}, unconventional power-scaling laws~\cite{tonkaev2024even,tonkaev2025unconventional}, controlling the polarization state and spin of harmonics \cite{jalil2022controlling, zhang2026tailoring}, wavefront shaping and focusing harmonics to a nanoscale spot \cite{korobenko2022situ}, and emerging hybrid platforms combining dielectric resonators with novel nonlinear materials including lithium niobate, halide perovskites, and van der Waals crystals~\cite{liu2025high,tonkaev2023observation,sakib2026polariton}. These developments demonstrate that resonant nanophotonic structures not only enhance strong-field interactions but also introduce entirely new mechanisms for manipulating HHG through resonance engineering, symmetry control, and material design.

In this review, we summarise the rapidly developing field of metaphotonics for HHG. We first provide a brief overview of HHG in gases, liquids, and solids before discussing the emergence of resonant sub-wavelength dielectric resonators and metasurfaces as efficient platforms for strong-field nonlinear optics. We then review recent developments based on high-$Q$ resonances, including bound states in the continuum and guided-mode resonances, highlighting how they enable efficient HHG, symmetry-controlled nonlinear processes, and unconventional nonlinear dynamics. Finally, we discuss generalized concepts based on symmetry engineering, novel material platforms, and hybrid photonic architectures that are expected to shape the next generation of compact high-harmonic light sources and integrated extreme nonlinear photonic devices.

\section{HHG: From gases to liquids and solids}

\begin{figure*}[!htbp]
    \begin{center}
    \includegraphics[width=1.0\linewidth]{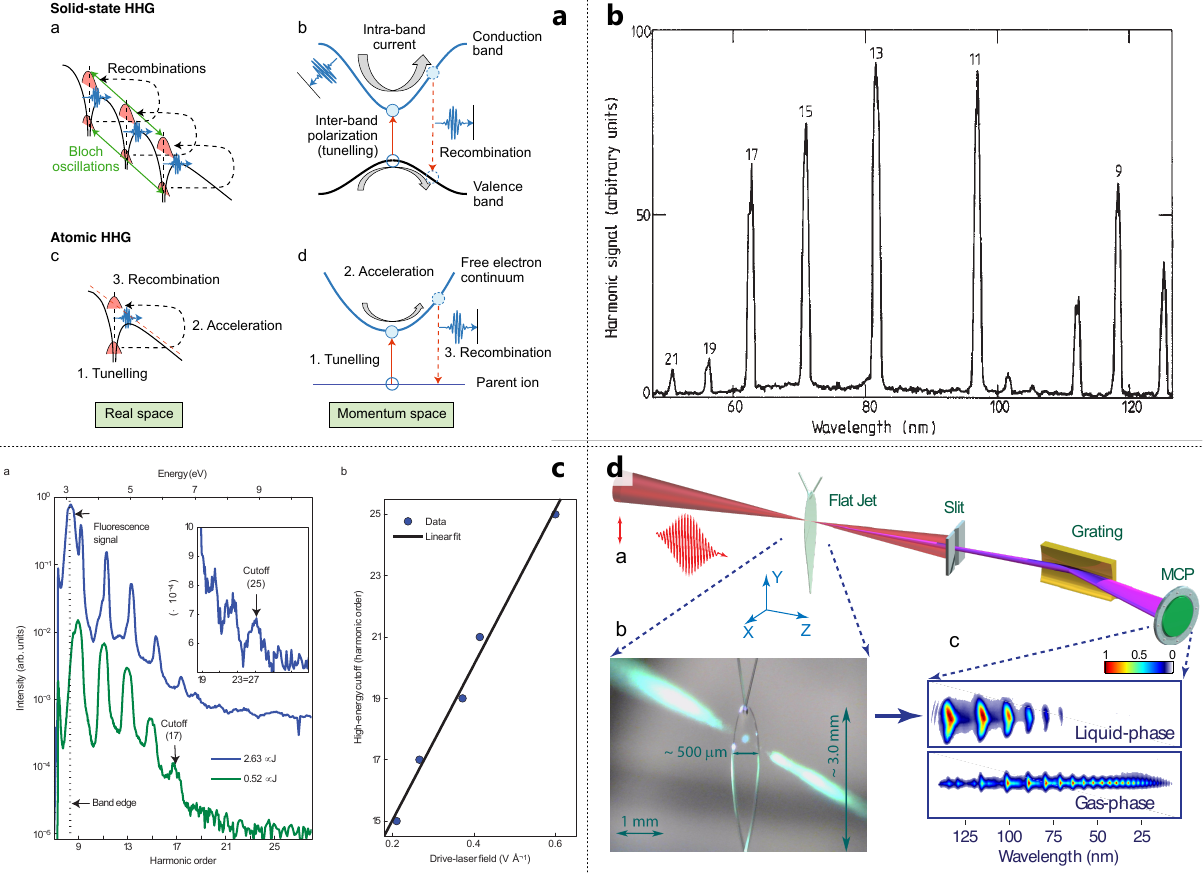}
    \caption{ \textbf{High harmonic generation: from gases to liquids and solids.} (a) Microscopic mechanisms of HHG in gas and solid, represented in real-space and momentum/reciprocal-space picture. Figure adapted from~\cite{ghimire2019high} \copyright 2019 Nature Springer. (b) The first experimental demonstration of non-perturbative HHG using xenon gas, showing up to 21st-order harmonics. Figure adapted from~\cite{ferray1988multiple} \copyright 1988 IOP Publishing Ltd. (c) The first experimental demonstration of HHG from bulk ZnO crystal, and the field strength dependence of its cutoff. Figure adapted from~\cite{ghimire2011observation} \copyright 2011 Nature Springer. (d) The first experimental demonstration of non-perturbative HHG from liquid water, with the detailed sketch of experimental apparatus, and the photograph of the laser-liquid interaction. Figure adapted from~\cite{luu2018extreme} \copyright 2018 Nature Springer. }
    \label{general}
    \end{center}
\end{figure*}

Until the work by Ferray et al. was published in 1988~\cite{ferray1988multiple}, the field of harmonic generation had been primarily evolving toward obtaining shorter and shorter wavelengths via harmonic generation excited by a short-wavelength laser field. However, non-perturbative high harmonics with a relatively broad plateau, i.e., high-conversion-efficiency region, could be generated from 11$^{th}$ to 17$^{th}$ harmonics~\cite{EarlyHHGGas_McPherson:87}. The sharp decrease of harmonics' efficiency with increasing the harmonic order limited the shortest wavelength that could be acquired. Ferray et al. took a different route, applied a longer-wavelength laser field, reaching the 33$^{rd}$-order harmonic corresponding to 32.2 nm from argon gas. Moreover, as shown in Fig.~\ref{general}\textbf{b} for argon, the detected harmonic spectrum displays a plateau-like structure from 9$^{th}$- to 17$^{th}$-order for the first time, followed by the sharp decrease in the higher-order region. This defines the typical harmonic spectrum in a gas: with harmonic order increasing, the harmonic spectrum shows three regions -- a steep decrease region, a plateau region and a sudden cutoff region~\cite{HHGGas_EarlyWork_XFLi1989}.

The theoretical foundations of strong-field physics have been established by Keldysh, in what we called 'Bible of Strong Field Physics'~\cite{KeldyshParameter_Keldysh1965}, only a few years later after the invention of laser in 1960~\cite{Maiman1960}. In this paper, the landmark 'Keldysh parameter' and its related theory quantitatively distinguish between multiphoton and tunnelling ionisation, providing the theoretical paradigm spanning the strong-field processes in gas and solid. After the experimental demonstrations of the high harmonics, the idea of the recollision was proposed by Kuchiev in 1987\cite{Kuchiev1987AtomicAntenna}. Kulander and Schafer\cite{Kulander1991,Kulander1991a} did a series of theoretical work on the theoretical calculations explaining the microscopic and macroscopic processes. Later in 1993, a simple-man picture was proposed by Corkum~\cite{HHGGas_3stepModel_PBCorkum1993}, intuitively describing the microscopic mechanism under semiclassical framework. Fig. \ref{general}\textbf{a} c presents the three-step re-collision model in a real-space picture, which has been widely applied to understand the HHG process in gases. In the tunnelling step, an electron becomes more likely to tunnel through the strong-laser-field-distorted atomic potential. Under the strong field approximation~\cite{HHGGas_SFA_Lewenstein1994}, which assumes the laser field dominates the motion of the electron, and the effect of the atomic potential can be ignored, the electron in continuum states is accelerated and acquires kinetic energy from the laser field. Until the electric field reverses its sign, the 'flying' electrons are dragged back, some of them return to and recombine with their parent ions, release the energy accumulated during acceleration as high harmonic radiation. In addition, the birth and return time of the 'flying' electrons, which are usually seen as the electron tunnelling time and recombining time, defining the short and long trajectories in real space~\cite{HHGGas_TrajectoryModel_PBalcou1997}, are imprinted into the high harmonic radiation, determining the intrinsic chirp of the generated attosecond pulses.

After a flourishing decade of seminal HHG observations in the 1990s, the first experimental realisation of an attosecond pulse train was reported~\cite{HHGGas_1stATPG_PMPaul2001}. From the generation of up to the 19$^{th}$-order harmonics in argon gas driven by 800-nm, 40-fs laser pulses, an attosecond pulse train was recorded, featuring a 250-as pulse duration and a 1.35-fs interval between adjacent pulses. This result opened up the field of extreme ultraviolet (EUV) attosecond light sources via gas-phase HHG. Since then, rapid progress has been made in this field towards the shortest pulse duration~\cite{HHGGas_ATPG_53as_Li2017,HHGGas_ATPG_43as_Gaumnitz2017,Han2026_18as}, and the highest photon energy~\cite{HHGGas_ATPG_1.6keV_TPopmintchev2012} -- an experimental roadmap of next-generation attosecond spectroscopy in EUV regime is gradually emerging.

The extensive understanding of HHG in gases has naturally paved the way for exploring HHG in solid-state materials, with crystalline systems emerging as a particularly attractive research focus. Compared with gases, the high atom density, periodicity, and engineered microscopic structures of solids bring not only new potential towards high-flux and high-energy attosecond light sources but also a fundamentally different understanding of attosecond electron dynamics~\cite{ghimire2011observation}. Transitioning from visible/near-infrared to mid-infrared laser sources, driving electrons farther from the Brillouin zone in solids, led to significant experimental breakthroughs in HHG for accessing non-perturbative strong-field physics, enabling electronic trajectory engineering, and serving as a spectroscopic probe of electronic structure. Starting from the first experimental demonstration of non-perturbative HHG from solids (shown in Fig. \ref{general}\textbf{c} a) reporting up to the 25$^{th}$-order harmonics~\cite{ghimire2011observation}, the subsequent works exhibit great promise across a wide range of applications. Consistent with prediction~\cite{ghimire2011observation}, HHG from solids has been shown to encode band-structure information, prompting Luu et al. to first propose a semiclassical framework of using spatial harmonic components to construct the 'basis set' of band dispersion, finally realising the reconstruction of materials' band structure. Under this framework, the ultrafast laser pulse plays the role of a 'probe', mapping the band structure at different momentum values in reciprocal space by driving electron Bloch oscillations across the Brillouin zone~\cite{HHGSolid_SiO2EUV_Luu2015}. Subsequently, the projection from orientation-dependent HHG spectra to the crystal structure and its symmetry was first established by correlating the enhancement and suppression of harmonic flux from MgO with the electron's transversal aligning and missing atomic sites~\cite{HHGSolid_CrystalSymmetryProbe_You2017}. In a related context, Berry curvature, one of the strong-symmetry-related geometric properties of materials, has been reported to be successfully retrieved through HHG in a $\rm{SiO_2}$ crystal~\cite{HHGSolid_BerryCurvExtra_Luu2018, uzan2024observation}, showing the potential of HHG as a powerful tool to study strongly-correlated systems, topological systems and valleytronic materials. In addition to the electronic structures, the lattice dynamics was theoretically predicted~\cite{HHGSolid_PhononDyn_ONeufeld2022} and later experimentally proven~\cite{HHGSolid_PhononDyn_Zhang2024} to be measurable by time-resolved high harmonic spectroscopy, with high time resolution and extreme sensitivity directly in energy domain, respectively. Moreover, HHG has also been reported to be a 'microscope', imaging the valence electron potential and density in real space with a spatial resolution reaching 26 pm, approximately the Bohr radius of atomic hydrogen~\cite{HHGSolid_ValenceElectronPicoscopy_Lakhotia2020}. Benefiting from the rapid development of condensed matter physics and material science, the application of solid HHG in this field is expected to hold even broader prospects beyond the above mentioned wide-ranging existing applications, offering more avenues for discovering and studying new physics.

HHG in gases has been commonly described semi-classically by three-step re-collision model, whereas HHG in solids was shown to require more thorough microscopic description. Recently, a unified perspective has been proposed to describe the phenomenon, by introducing vacuum electric field fluctuations for a full-quantum description \cite{thorpe2025high}. Figure \ref{general}\textbf{a}: a - b schematically show the mechanisms associated with solid HHG in a real-space picture and a momentum-space picture, respectively, with Fig. \ref{general}\textbf{a}: c - d showing a counterpart in gas HHG. The difference between these two mechanisms mainly originates from the high atom density and the periodic atom arrangement in solid. In real-space picture, the high atom density leads to de-localised electron states, via superposition of atomic orbitals. Consequently, when the electron is excited from the ground state to excited state, it cannot be treated as a free electron being driven solely by laser field, instead, it remains confined by the periodic lattice potential, influenced by laser field. Similarly, the subsequent 'recombination' step cannot be considered as the recombination with electron's parent ion, rather, the probability of electron being trapped by the potential well of other ions is increased due to the same reason. In momentum space, based on band theory, the model to describe solid HHG process is more intuitive, where the discrete levels in gas (shown in Fig. \ref{general}\textbf{a}: d) evolve to dispersive conduction and valence bands in solid (shown in Fig. \ref{general}\textbf{a}: b). The interband transition and intraband Bloch oscillation are the main contributions to HHG, which provides a basic description of solid HHG in strong laser field with moderate strength. It is worth noting, however, at the extreme intense field strength, laser field drastically modifies the lattice potential and the band structure. In this limit, the suppression of lattice potential converts the band-to-band-transition picture to a laser-field-dominated, quasi-free-electron-like picture~\cite{HHGSolid_ValenceElectronPicoscopy_Lakhotia2020}.

During the last two decades, significant progress in modelling of HHG in transparent solids has been achieved. As soon as it was realised that the electrons in a crystal cannot be treated as continuum free electrons, three major approaches emerged to predict HHG \cite{yu2019high}: time-dependent Shr\"{o}dinger equation (TDSE) \cite{plaja1992high, wu2015high}, semiconductor Bloch equations (SBE) \cite{haug1994quantum, golde2008high}, and time-dependent density-functional theory (TDDFT) \cite{runge1984density}. None of these approaches is exhaustive: TDSE depends on the choice of Hamiltonian, SBE on independent particle approximation, and TDDFT on the exchange-correlation potential under high nonequilibrium. Altogether they provide different insights into the involved complex physics. SBE has become a quantitative workhorse to relate material properties and laser irradiation conditions directly to HHG spectra, evolving from two-band conceptual models \cite{meier1994coherent, golde2008high} to multiband models \cite{schubert2014sub,Luu2016,huttner2017ultrahigh}, incorporating transition dipole phases \cite{jiang2017effect}, Berry connection \cite{yue2020structure, li2019phase}, three-dimensional full-Brillouin-zone \cite{gu2022full, parks2025full}, and many-body Coulomb interactions \cite{parks2026many}. Recently, first-principles simulations using TDDFT have become computationally feasible to describe nonlinear electron dynamics on attosecond timescales, naturally including many bands and realistic crystalline electronic band structure \cite{freeman2022high}. Extending the study from point models to higher dimensions, a few recent works explored the role of nonlinear propagation effects in one-dimensional slabs, by applying the microscopic approaches coupled with Maxwell solvers \cite{kilen2020propagation, rudenko2022self, yamada2023propagation, kolesik2024propagation}.

Motivated by the pursuit of high-energy attosecond light sources, as introduced earlier, the determination of the high harmonic's cutoff energy has been an important subject from the outset. In fact, the cutoff energy determines the available attosecond bandwidth, while the spectral phase and temporal coherence of the plateau determine whether this bandwidth can be compressed into an attosecond pulse.  In the early stage, a simple formula was proposed to describe cutoff energy in the harmonic spectra for atoms and ions with linear dependence on the square of laser field strength, and the square of laser  wavelength~\cite{HHGCutoff_EarlyWork_Krause1992}. However, owing to the lower damage threshold of solids and competing interband and intraband contributions, the dependence of cutoff energy in solid HHG - unlike the case in gas - is not universal, depending not only on the electric field amplitude but also on electronic band structure. A prevailing perspective states that the cutoff energy undergoes a linear increase with field strength, supported by experimental observations (Fig. \ref{general}\textbf{a} c) and the corresponding simulations~\cite{ghimire2011observation,HHGSolid_SiO2EUV_Luu2015}. While this mainstream view focuses on the field strength dependence, peripheral discussions exist regarding the dependence on wavelength~\cite{wu2015high, liu2017wavelength, WavelengthDep_Cutoff_Wan2023}, effect of carrier-envelope phase and dispersion~\cite{HHGCutoff_CEP_Dispersion_Zhong2016}, and sensitivity to ellipticity of input laser field~\cite{HHGCutoff_Ellipticity_Tancogne-Dejean2017}, covering experimental and theoretical works.

Beyond the different dependence law of cutoff energy in solid HHG, benefiting from crystal's periodic structure, the symmetry-constrained HHG offers a new dimension to explore nonlinear optics, and deserves particular attention. From the very beginning of experimental studies, Ghimire et al. have noticed the features of crystal symmetry-dependent harmonic flux change~\cite{ghimire2011observation}. Subsequently, the first detailed discussion came in 2017, establishing a real-space picture that the high harmonics are enhanced when the electron trajectories connect nearby atomic sites, covering from 13th to 21st harmonics~\cite{HHGSolid_CrystalSymmetryProbe_You2017}. The same phenomenon was addressed later that year from a complementary perspective, with Wu et al. proposing a momentum-space interpretation, which attributes orientation dependence of high harmonics to the embedded symmetry properties of band structure~\cite{HHGSolid_SymmCons_TDSE_MWu2017}.

Generation of even harmonics is another important symmetry-related phenomenon, which was primarily associated with structural broken inversion symmetry~\cite{ghimire2011observation}, without indicating specific microscopic mechanism or quantity responsible for symmetry breaking. A few years later, non-perturbative quantum interference between direct and indirect excitation pathways was proposed to interpret even HHG \cite{hohenleutner2015real}. The theoretical description was further completed by introducing transition dipole phases \cite{jiang2017effect} and Berry curvature contributions through asymmetric carrier motion and intraband polarisation \cite{HHGSolid_BerryCurvExtra_Luu2018}. It was later understood that the HHG sources are distributed throughout the full-Brillouin-zone and have different phases, often interacting destructively, thus, restricting the calculation to a high-symmetry line or a small $k-$space region with large individual contributions, as implemented before, can give qualitatively incorrect symmetry and even-order harmonics amplitudes \cite{gu2022full}. Furthermore, due to the reduced dimensionality in two-dimensional materials, the broken symmetry induced effects are more significant, introducing new possibilities for exploring symmetry constrained HHG within non-perturbative regime. For instance, recent works have been devoted to separate the surface and bulk contributions in harmonic generation \cite{li2026disentangling, jiang2023distinguishing}. Surface nonlocal and high-order magnetic and electric quadrupole contributions were proposed to complete the HHG picture \cite{walser2000high}, to be comparable to dipolar emission at high photon energies \cite{gorlach2020quantum}, and to inversion-symmetry breaking at the surface of centrosymmetric materials such as silicon \cite{sipe1987phenomenological, smirnova2018, hallman2025high}. These mechanisms for even-order harmonics generation in solids have been recently addressed by ab initio TDDFT simulations \cite{tancogne2016ab,  jensen2025beyond}. Alternatively, hydrodynamic-Maxwell approaches, based on nonlinear Maxwell equations with nonlocal terms, corresponding to magnetic, convection and high-order electric terms, were applied to describe macroscopic spatially nonlocal nonlinear optical response. Initially proposed in the context of nonlinear plasmonics \cite{sipe1980analysis, ginzburg2015nonperturbative, krasavin2018free}, such approaches were further adopted to simulate odd- and even-order harmonic generation in conductive oxides and semiconductors \cite{rodriguez2019harmonic, scalora2020electrodynamics, tonkaev2024even, hallman2025high}. They can include also the intraband current for conduction band electrons, with the electron density evolving via Keldysh photo- and avalanche ionisation and recombination \cite{rudenko2018photogenerated, sinev2021observation, zalogina2023high}. The inversion symmetry can be also broken dynamically by spatially inhomogeneous electric fields \cite{ciappina2012high, yavuz2012generation, du2016enhanced, rudenko2018photogenerated} and asymmetric surface charge accumulation \cite{jalil2023spectroscopic}, suggesting an additional source for even HHG, not previously included in semi-classical SBE \cite{zhang2025spatially}.

Recent experiments suggest that metallic nitrides \cite{korobenko2021high} and noble metals \cite{gholam2025high} can also act as efficient sources for HHG. In the latest work, the non-perturbative HHG up to the 13th  order were found to originate from coherent electron dynamics in the silver crystal lattice within the material's penetration depth and applying fluence below the multi-shot modification threshold \cite{gholam2025high}. In contrast to transparent solids, metals  initially have  a high density of conduction band electrons to generate strong currents, the carriers are confined strongly to the interface and may have very different dominant relaxation pathways, such as electron-electron, electron-phonon and surface scattering rather than de-phasing. The nonlinearity can originate from Fermi-surface intraband dynamics but also interband transitions between non-parabolic bands, similar to dielectrics and semiconductors. In addition, the harmonic spectra can have signatures from electron photo- and field emission and generated metal plasma at high intensities, which have different harmonic polarisation dependencies, making it possible to differentiate between the contributions. Finally, it remains unknown whether surface contributions, magnetic dipole and electric quadrupole contributions, so far proposed to dominate in perturbative SHG from metallic thin films and nanoparticles \cite{dadap2004theory, wang2009surface, kujala2007multipole}, can equally contribute to non-perturbative even HHG.

A nascent area of HHG research is its extension to liquids, which has only recently begun to be studied. Liquids combine a high atomic density—comparable to that of crystalline solids—with a lack of long-range order~\cite{HHGLiquid_BookChap_Worner2024}, a characteristic they share with gases and amorphous systems. These characteristics have already made liquid an attractive system in the HHG field. Moreover, as a fertile ground for liquid-phase chemistry and biology, liquid exhibits charge and energy transfer time scales predominantly in the femtosecond to attosecond regime, which calls for probes combining ultrashort temporal resolution, high spatial resolution, and in situ operation -- HHG, which has been investigated for a few decades, stands out as an excellent choice. However, despite the wealth of HHG expertise in gas and solids, the observation of real HHG from liquid has proven more challenging. In the early 2000s, only incoherent plasma luminescence was observed from laser-interaction-expanded water droplets, although reaching around 40 nm regime ~\cite{HHGLiquid_EarlyWork_ExpandedDroplets_Flettner2003}. Later, the coherent, visible harmonic emission was observed in $\sim$ 150 $\rm{\mu}$m thick bulk liquid sample, but still fell in perturbative regime~\cite{HHGLiquid_EarlyWork_BulkLiquid_DiChiara2009}. The realization of detecting the non-perturbative, EUV HHG from pure bulk liquid sample -- not an expanded liquid, not a mixture of the surrounding gas and inner liquid, remains a major technical obstacle. It was not until 2018, with the advent of liquid flat-microjet technology, that EUV HHG was observed from liquid water and several alcohols~\cite{HHGLiquid_1stWork_Luu2018}. Significantly, HHG from the pure liquid sample and the surrounding gas at the interaction area was spatially separated owing to the unique wedge-like design of the liquid jet. As shown in Fig. \ref{general}\textbf{d} c, the cutoff energy shows linear dependence on the laser field strength, reaches beyond 20 eV, heralding the arrival of the real coherent EUV radiation stage in liquid HHG field.

The exploration towards the highest photon energy of liquid HHG never stops. Although some subsequent work reports the cutoff energy staying at the same level as the first liquid HHG work~\cite{HHGLiquid_Cutoff_Svoboda2021,HHGLiquid_MechExplo_Cluster_Mondal2023,HHGLiquid_Cutoff_PulDur_Mondal2023}, a new maximum cutoff energy at around 50 eV refreshes the record ~\cite{HHGLiquid_Cutoff_AOliver2023}, stimulating new thinking about the underlying mechanism that the lack of long-range order is the main limitation of extending the cutoff energy in liquid HHG. Furthermore, the dependence law of cutoff in liquid HHG has also been investigated from the first experimental demonstration work~\cite{HHGLiquid_1stWork_Luu2018}, exhibiting the similar linear dependence of laser field strength to solid HHG. In parallel, the dependence of ellipticity~\cite{HHGLiquid_MechExplo_Cluster_Mondal2023}, the independence of pulse duration~\cite{HHGLiquid_Cutoff_PulDur_Mondal2023} on cutoff energy, and a linear dependence on the square of laser field strength~\cite{HHGLiquid_Cutoff_AOliver2023}, instead, reveal the similarity of mechanisms to gas HHG, manifesting the crucial impact of electron scattering. The long-standing consensus that the microscopic structure of liquid has the features of both gas and solid has been recently challenged by observation of multi-plateau HHG driven by off-site recombination with neighbouring molecules \cite{mondal2026multi}. The involved mechanism, different in nature compared to both gas and solid phases, was supported by ab initio TDDFT simulations, paving the route of ultrafast spectroscopy of electron dynamics in solution.

While studies of HHG from gases have reached the maturity since their start four decades ago, studies of HHG in condensed matter had just begun by more than a decade, with an overwhelming number of directions and potential applications. Among those interesting directions, combination of HHG with metamaterials has started taken shape.

\section{Subwavelength resonators}

The emergence of HHG in solids has stimulated intense efforts to manipulate strong-field light-matter interactions using nanophotonic structures. One of the earliest approaches relies on subwavelength resonators, which enable strong localization and enhancement of optical fields. Such local electric-field enhancement is crucial for driving highly nonlinear optical processes. Initial attempts to enhance HHG employed plasmonic nanoantennas capable of concentrating optical energy into nanoscale hotspots. However, because metals possess relatively weak intrinsic nonlinearities, these hotspots were typically filled with gases, where HHG was generated~\cite{kim2008high,sivis2012nanostructure,park2013generation,sivis2013extreme}. Later, this approach was extended to solid-state nonlinear media~\cite{vampa2017plasmon,imasaka2018antenna,jalil2023spectroscopic} and to nanostructured waveguide geometries such as tapered cones~\cite{han2016high, franz2019all}. Despite these advances, the practical implementation of plasmonic platforms remains limited by Ohmic losses, which lead to optical heating and low damage thresholds under intense excitation. Furthermore, efficient HHG often requires the incorporation of additional nonlinear materials into the plasmonic hotspots. Without the plasmonic counterparts, a tapered sapphire cone can still support HHG but the efficiency drops significantly \cite{han2022high}.

\begin{figure*}[!htbp]
    \begin{center}
    \includegraphics[width=0.99\linewidth]{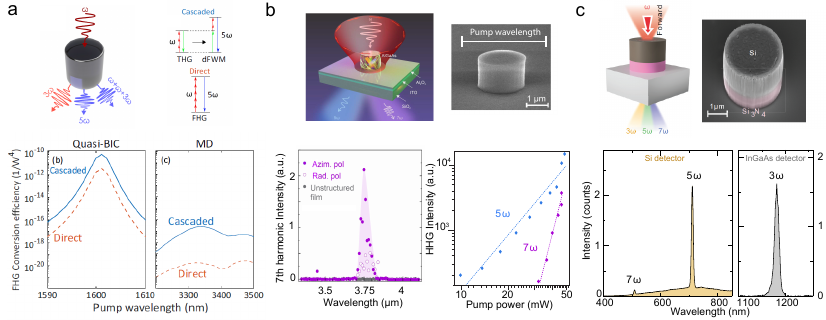}
    \caption{ \textbf{High-harmonic generation from subwavelength resonators.} (a) Fifth-harmonic generation from a single resonator including cascading effects. Top: Schematic illustration of direct and cascaded fifth harmonic generation. Bottom: Calculated fifth-harmonic conversion efficiency in the vicinity of a quasi-bound state in the continuum and a magnetic dipole resonance for direct and cascaded processes. Figure adapted with permission from~\cite{carletti2019high} \copyright 2019 American Physical Society. (b) AlGaAs nanoresonator supporting resonances in the mid-infrared. Top: schematic illustration of high-harmonic generation from a subwavelength resonator under azimuthally polarised excitation together with an SEM image of the fabricated structure. Bottom: Seventh-harmonic intensity under azimuthally and radially polarized excitation, compared with that from an unstructured film, and power dependencies of the fifth- and seventh-harmonic signals under resonant excitation. Figure adapted with permission from~\cite{zalogina2023high} \copyright 2023 AAAS. (c) Si/Si$_3$N$_4$ resonator supporting resonances in the mid-infrared~\cite{mathew2026asymmetric}. Top: schematic illustration of the asymmetric high-harmonic generation from a subwavelength resonator together with an SEM image. Bottom: high-harmonic spectra under resonant excitation.}
    \label{fig:sub_res}
    \end{center}
\end{figure*}

These limitations motivated a transition toward all-dielectric resonators supporting Mie-type optical modes~\cite{kuznetsov2012magnetic, kuznetsov2016optically, kivshar2017meta}. High-index dielectric nanoparticles exhibit both electric and magnetic resonances while maintaining significantly lower dissipative losses than their plasmonic counterparts. Importantly, dielectric materials possess intrinsic optical nonlinearities and support strong field enhancement directly within the resonator volume, enabling efficient nonlinear light–matter interactions without the need for additional nonlinear media. As a result, harmonic generation can occur directly inside the nanostructure.

Early studies of Mie-resonant nanostructures demonstrated dramatic enhancements of second- and third-harmonic generation efficiencies, establishing subwavelength resonators as a powerful platform for nonlinear nanophotonics~\cite{grinblat2017degenerate, makarov2017efficient, timpu2017second, timpu2019lithium, popkova2022nonlinear}. The application of dielectric resonators to HHG represents a natural extension of these concepts into the strong-field regime. Recent advances have further expanded this concept through the introduction of resonances with increasingly high quality factors. In particular, quasi-bound states in the continuum (quasi-BICs) provide exceptional field confinement and extended photon lifetimes. These resonances have enabled significant enhancement of low-order harmonic generation in subwavelength resonators~\cite{koshelev2020subwavelength}.

In a theoretical work, Carletti \textit{et al.}~\cite{carletti2019high} proposed a novel approach for enhancing HHG in a single subwavelength dielectric resonator by exploiting optical bound states in the continuum. The authors considered a high-index dielectric nanoparticle placed on an epsilon-near-zero substrate (Fig. ~\ref{fig:sub_res}a), where the coupling between resonator modes and substrate-induced image modes gives rise to a supercavity state with an ultrahigh quality factor. Using a nonlinear coupled-mode framework, they demonstrated that the strong field confinement associated with the BIC can dramatically enhance fifth-harmonic generation through both direct and cascaded nonlinear processes occurring within a single nanoresonator. The study predicted orders-of-magnitude enhancement of harmonic yields compared with conventional Mie resonances and established BIC-driven resonators as a promising platform for nanoscale strong-field nonlinear optics.

Building directly on the concept of BIC-enhanced nonlinear response in individual subwavelength resonators, Zalogina \textit{et al.}~\cite{zalogina2023high} provided an experimental realisation of this idea. The authors demonstrated that quasi-BIC resonances in a single AlGaAs dielectric nanoresonator can strongly enhance HHG. The resonator supports a quasi-BIC mode in the mid-infrared spectral range around 3.7~$\mu$m (Fig. ~\ref{fig:sub_res}b) and is excited using a tightly focused azimuthally polarized beam, enabling efficient coupling to the high-Q mode and strong field confinement. The authors experimentally observed harmonic generation up to the seventh order, while no measurable harmonic signal was detected from an unstructured AlGaAs film under the same excitation conditions. These results provided clear experimental evidence that quasi-BIC resonances in individual dielectric nanoresonators can dramatically enhance strong-field nonlinear optical processes and enable efficient HHG from deeply subwavelength volumes.

Building on recent advances in BIC- and Mie-resonant subwavelength structures for enhanced nonlinear optics, Jangid \textit{et al.} recently demonstrated a complementary approach for controlling HHG through structural asymmetry in individual nanoresonators. They experimentally showed that bilayer Si/Si$_3$N$_4$ cylindrical resonators supporting coupled electric and magnetic Mie modes exhibit pronounced asymmetry in nonlinear emission, resulting in strongly different third-, fifth-, and seventh-harmonic generation efficiencies under forward and backward illumination (Fig. ~\ref{fig:sub_res}c). This asymmetry originates from direction-dependent magnetoelectric coupling between resonant modes, which modifies the multipolar composition and local field enhancement inside the resonator depending on the excitation direction. Measurements performed in the mid-infrared revealed a strong forward–backward contrast in harmonic intensities that could be tuned through both the resonator diameter and pump wavelength. This work extends the concept of nonlinear control in single resonators beyond resonance enhancement alone, introducing directional engineering of strong-field nonlinear processes as an additional degree of freedom for compact HHG sources.

In contrast to HHG from bulk samples, the microscopic mechanisms in the origin of high-order harmonics can be additionally affected by the field and plasma-induced inhomogeneity in subwavelength nanostructures \cite{taghinejad2020transient, an2021efficient, jalil2023spectroscopic}. In the latest work, the symmetry was broken by inhomogeneous fields induced by gold nano-triangles on a silicon substrate and the sixth harmonic was detected from centrosymmetric semiconductor \cite{jalil2023spectroscopic}. In perspective, ultrafast symmetry breaking in nanostructures can be studied independently of material crystallographic orientation.

\section{Metasurfaces}

The transition from individual subwavelength resonators to metasurfaces marks a conceptual shift in metaphotonics for HHG, from localized field enhancement to engineered collective optical responses. While single nanoresonators provide extreme spatial confinement of the electromagnetic field, their practical application is limited by the small active nonlinear volume and the lack of spatial phase control over the emitted nonlinear signal. Metasurfaces, planar arrays of subwavelength scatterers, overcome these limitations by enabling control over the macroscopic phase, amplitude, and polarization of the generated harmonics, while simultaneously increasing the overall nonlinear signal through a larger effective interaction volume.

Early developments in this direction were driven by plasmonic metasurfaces composed of metallic nanoantenna arrays. These systems employed strong local field enhancement associated with localized surface plasmon resonances~\cite{kauranen2012nonlinear,butet2015optical}. Since metals exhibit relatively weak intrinsic nonlinearities and significant losses, plasmonic nonlinear metasurfaces typically rely on surrounding nonlinear media or embedded dielectrics, where nanoantennas act as local field-enhancing hotspots. In this context, silicon and other nonlinear materials have been widely employed to demonstrate enhanced harmonic generation in plasmonic metasurfaces~\cite{vampa2017plasmon, wan2025high}. This approach also enables control over the polarization and phase of the generated harmonics~\cite{jalil2022controlling}, including the generation of circularly polarized harmonic emission facilitated by plasmonic enhancement in silicon-based systems~\cite{ren2024plasmon}.

Plasmonic metasurfaces inherit intrinsic limitations of metallic systems, most notably Ohmic losses and limited damage thresholds under strong-field excitation. These constraints have motivated the rapid development of all-dielectric metasurfaces. Dielectric metasurfaces not only significantly reduce dissipative losses but can also be engineered from materials with strong intrinsic nonlinearities, enabling more efficient and versatile control over nonlinear optical processes. One of the first demonstrations of enhanced HHG in this platform was the observation of fifth-harmonic generation enhancement in argon driven by an all-dielectric metasurface, with a reported enhancement factor of up to 45~\cite{ginsberg2020enhanced}.

\begin{figure*}[!htbp]
\begin{center}
\includegraphics[width=0.99\linewidth]{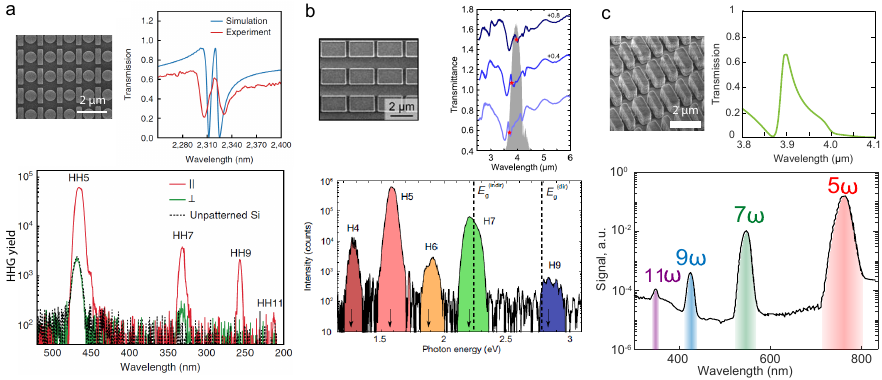}
\caption{\textbf{High-Harmonic Generation from resonant metasurfaces.} (a)  Silicon metasurface supporting a Fano resonance in near-infrared. Top: SEM image and linear transmission showing the resonance position. Bottom: HHG spectra from the metasurface pumped with two orthogonal linear polarisations, compared with an unpatterned silicon film. Figure adapted with permission from~\cite{liu2018enhanced} \copyright 2018 Springer Nature. (b) Ultrathin resonant gallium phosphide metasurface for even- and odd-order harmonics generation. Top: SEM image and normal-incidence transmission spectra of three samples with different dielectric resonator sizes exhibiting resonances in mid-infrared. Bottom: HHG spectra from the metasurface under resonant mid-infrared excitation. Figure adapted with permission from~\cite{shcherbakov2021generation} \copyright 2021 Springer Nature. (c) Silicon metasurface supporting a bound state in the continuum in the mid-infrared. Top: SEM image and calculated linear transmission spectrum showing resonance. Bottom: HHG spectra from the metasurface under resonant excitation. Figure adapted with permission from~\cite{zograf2022high} \copyright 2022 ACS.}
\label{fig:metasurfaces}
\end{center}
\end{figure*}

A major breakthrough in solid-state HHG was achieved by Liu et al., who demonstrated resonantly enhanced non-perturbative HHG from an all-dielectric silicon metasurface supporting a sharp Fano resonance arising from a classical analogue of electromagnetically induced transparency~\cite{liu2018enhanced}. The metasurface consisted of periodically arranged silicon bar resonators coupled to disk resonators, producing strong field confinement while maintaining the high damage threshold characteristic of dielectric materials (Fig. ~\ref{fig:metasurfaces}a). Compared with an unpatterned silicon film, the metasurface enhanced harmonic emission by more than two orders of magnitude, clearly demonstrating that resonant nanophotonic engineering can dramatically modify strong field light-matter interactions. Moreover, the harmonic yield exhibited pronounced polarization and wavelength selectivity, with efficient generation occurring only when the excitation was aligned with the resonant mode and tuned to the Fano resonance. This work marked the first demonstration that all-dielectric metasurfaces can serve as efficient platforms for engineering strong-field phenomena, laying the foundation for subsequent developments based on high-$Q$ resonances and bound states in the continuum for HHG.

Building on the success of silicon metasurfaces, Shcherbakov et al. extended the concept to resonant metasurfaces fabricated from gallium phosphide (GaP), a wide-bandgap, non-centrosymmetric semiconductor that combines high refractive index with low absorption in the visible spectral range \cite{shcherbakov2021generation}. The dielectric metasurface consisted of arrays of resonant GaP nanoantennas supporting electric dipole modes in the mid-infrared, providing moderate local field enhancement while avoiding the strong free-carrier absorption that limits silicon at high pump intensities (Fig. ~\ref{fig:metasurfaces}b). Owing to the broken inversion symmetry of GaP, the metasurface generated both even- and odd-order harmonics, with harmonic orders up to the ninth observed under resonant excitation. Importantly, the use of single-shot excitation enabled the authors to reach significantly higher peak intensities without cumulative laser-induced damage, allowing them to directly probe the transition from perturbative to non-perturbative high-harmonic generation. This work highlighted material selection as an essential degree of freedom alongside resonant design, establishing a robust platform for strong-field nanophotonics.

A further advance in resonant dielectric metasurfaces was achieved through the introduction of quasi-BICs, which enable significantly stronger field confinement and higher Q-factors than conventional Mie or Fano resonances. Zograf et al. employed symmetry-broken silicon metasurfaces supporting quasi-BIC resonances to demonstrate efficient HHG up to the 11th order under mid-infrared excitation \cite{zograf2022high} (Fig. ~\ref{fig:metasurfaces}c). By systematically investigating the harmonic spectra, the authors observed a transition from perturbative to non-perturbative nonlinear response, where lower-order harmonics followed conventional power-law scaling while higher-order harmonics exhibited signatures of strong-field electron dynamics. The extreme field enhancement provided by the quasi-BIC resonance enabled efficient harmonic generation without sacrificing the high damage threshold of dielectric materials, illustrating the unique advantages of high-$Q$ resonances for strong-field nanophotonics.

A complementary demonstration of resonant enhancement of HHG was reported by Peterka et al., who investigated amorphous silicon metasurfaces supporting broadband magnetic dipole resonances in periodic nanoantenna arrays \cite{peterka2023high}. In contrast to high-Q quasi-BIC approaches, this work exploited a lower-Q but spectrally broad magnetic resonance to enhance nonlinear emission under ultrashort femtosecond excitation. The metasurface, composed of periodically arranged amorphous silicon disks on a dielectric substrate, provided strong local magnetic-field confinement associated with Mie-type resonances, leading to an approximately twenty-fold enhancement of high-harmonic yield compared to unstructured films. In parallel, a range of theoretical works have explored symmetry-protected BICs, polarization-selective resonances, and dual-Fano configurations for optimizing HHG in metasurfaces~\cite{xiao2022robust,xiao2022polarization,sun2024enhanced,liu2025high,khakpour2024enhanced}.

Building on these developments, free-standing dielectric membrane metasurfaces have emerged as an effective platform for enhancing HHG by eliminating substrate-induced losses and further increasing optical confinement. Low-Q collective resonances~\cite {sartorello2024nonlinear} and Fabry Perot resonance~\cite{hallman2025high} have demonstrated significant enhancement of fifth-harmonic generation near resonance conditions. More recently, high-Q free-standing membrane metasurfaces have enabled enhancement of high-harmonic emission by up to three orders of magnitude~\cite{tonkaev2025unconventional}. In addition, the absence of a substrate introduces additional symmetry degrees of freedom, enabling advanced symmetry-engineering strategies for nonlinear optical control. These effects have recently been extended to chiral HHG, experimentally demonstrated in lower-order nonlinear processes~\cite{tonkaev2025nonlinear, hariharan2026nonlinear}, highlighting the potential of membrane metasurfaces for full vectorial and helicity control of strong-field light-matter interactions.

High-harmonic emission can also be realised in reflection geometry, which could be beneficial to avoid propagation effects or to achieve phase matching, when the high harmonics are absorbed by the target \cite{xia2018nonlinear, vampa2018observation, heinrich2021chiral}.
From the application side, the challenge lies in in situ focusing of high harmonics at the nanoscale, that can be achieved, for instance, by designing Fresnel zone plate metalens at the target exit \cite{korobenko2022situ, bastani2025model}. Alternative designs have been proposed to manipulate extreme ultra-violet light at nanoscale using nanoholes and nanopillars \cite{ossiander2023extreme, liang2026metasurfaces}. Another direction that might be promising is designing metasurfaces for frequency mixing, with an optical frequency mixer realised by two pump laser beams so far for low-order harmonics \cite{liu2018all}, and extending sum-frequency mixing to high-order harmonics \cite{xiao2022polarization}.

Self-consistent modelling of nonlinear optical response of metasurfaces, extending to microscopic dimensions, still remains challenging. One of the issues is the incompatibility of computationally intensive and memory-consuming microscopic quantum methods, such as TDDFT, TDSE or full-Brillouin-zone resolved SBE, commonly applied to a point in space, with 3D propagation scales and metasurface dimensions. In fact, upon ultrafast irradiation, metasurfaces turn into many-emitter quantum sources, that can have different surface or bulk microscopic origins and can additionally interact locally or nonlocally, constructively or destructively with each other at the nanoscale. Intuitively, the problem can be sub-divided into pre-calculation of microscopic nonlinear currents, combining them to study an individual nonlinear response of each meta-atom and then extending to metasurfaces and further to far-field where the signal is measured. However, such sub-division neglects both ultrafast inter-coupled dynamics and spatial nonlocal effects that also matter in shaping the HHG spectra. On the other hand, while focusing mainly on these effects, strong approximations involving empirical models and simplified electronic band structure, have so far been applied to resolve the microscopic dynamic responses in each spatial point. The research field of ultrafast nonlinear metaphotonics would strongly benefit from advanced methods that would incorporate self-consistently microscopic electronic response and full-vector nonlinear propagation in metasurfaces.

\section{Generalised concepts and novel materials}

The rapid evolution of metaphotonic platforms for HHG has revealed that resonant field enhancement alone is insufficient to fully control strong-field nonlinear processes. Instead, recent developments point toward a more general framework in which the interplay between optical resonances, structural symmetry, and intrinsic material nonlinearity governs both the efficiency and the properties of harmonic emission~\cite{kang2022high,koshelev2024scattering,yi2025efficient}. The symmetry of individual meta-atoms, the lattice, and the underlying crystal orientation determines the allowed nonlinear optical transitions and the angular momentum of emitted harmonics through selection rules, enabling deterministic control over the order, polarization, and helicity of the generated radiation~\cite{chen2014symmetry,ren2024plasmon}. Beyond geometry, the emergence of novel nonlinear materials, including  transition metal dichalcogenides~\cite{yoshikawa2019interband}, halide perovskites~\cite{hirori2019high}, epsilon-near-zero materials~\cite{yang2019high}, GST~\cite{korolev2024tunable} and other strongly nonlinear platforms~\cite{long2023high, ghimire2011observation}, has further expanded the accessible parameter space, allowing material-specific electronic properties to be exploited alongside photonic design for efficient and tailored HHG.

\begin{figure*}[!htbp]
\begin{center}
\includegraphics[width=0.99\linewidth]{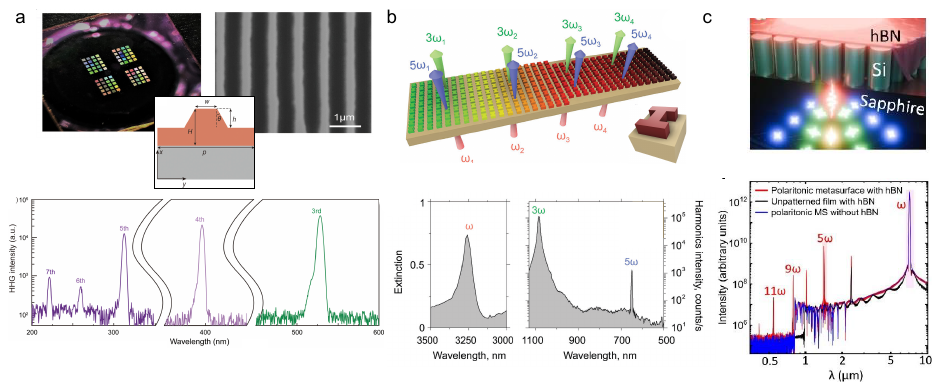}
\caption{\textbf{High-harmonic generation: novel concepts and materials}. (a) Lithium niobate metasurface supporting bound states in the continuum and guided-mode resonances in the near-infrared. Top: optical photograph and SEM image of the fabricated LiNbO$_3$ metasurface together with a schematic illustration of a metasurface unit cell. Bottom: Measured high-harmonic spectra from the metasurface under resonant excitation, exhibiting both even and odd harmonics up to the 7th order. Figure adapted with permission from~\cite{zhao2023efficient} \copyright 2023 Royal Society of Chemistry. (b) Germanium gradient metasurface with a bone-shaped meta-atom geometry and a gradually tunable resonance position. Top: schematic of the gradient metasurface for spectrally tunable enhancement of high-harmonic generation. Bottom: experimental linear and nonlinear spectra of the metasurface under resonant excitation, showing a fundamental resonance and its third and fifth optical harmonics. Figure adapted with permission from~\cite{jangid2024spectral} \copyright 2023 John Wiley $\&$ Sons. (c) Polaritonic metasurface–mediated high-harmonic generation in a hybrid structure~\cite{sakib2026polariton}. Top: Schematic illustration of a silicon polaritonic metasurface coupled with hexagonal boron nitride. Bottom: high-harmonic generation calculations for three different cases, suggesting that adding a thin layer of hexagonal boron nitride on top of a polaritonic metasurface can lead to a three orders of magnitude enhancement of fifth-harmonic generation.}
\label{fig:gener}
\end{center}
\end{figure*}

Lead halide perovskites exhibit exceptionally strong optical nonlinearities~\cite{hirori2019high} making them highly attractive for nonlinear metaphotonics. Recent work~\cite{tonkaev2023observation} demonstrates experimentally the first resonantly enhanced fifth-harmonic generation in a perovskite metasurface. The system consists of a 130 nm-thick perovskite slab patterned with a shallow grating, supporting highly nonlocal lattice resonances in the visible spectral range, specifically a bound state in the continuum and a guided-mode resonance. Importantly, these modes are engineered to be quasi-degenerate at the harmonic wavelength rather than at the pump, enabling efficient nonlinear conversion under mid-infrared excitation. The metasurface exhibits a broadband enhancement of fifth-harmonic emission (2750–3000 nm pump range), reaching up to two orders of magnitude compared to an unpatterned film, while third-harmonic generation remains largely unenhanced due to the absence of resonances at that wavelength. The enhancement is attributed to resonant field buildup and mode excitation of nonlocal lattice modes, with the broad spectral response arising from strong dispersion and k-space averaging of the resonances.

A further extension of material-driven nonlinear metaphotonics was demonstrated in lithium niobate (LiNbO$_3$) metasurfaces, where guided-mode resonances were used to enhance both second- and higher-order harmonic generation in a material platform with exceptionally large second-order nonlinear susceptibility and wide optical transparency window \cite{zhao2023efficient}. By structuring LiNbO$_3$ into subwavelength resonant nanostructures, the authors overcame the intrinsic phase-matching limitations of bulk crystals and achieved efficient harmonic generation up to the seventh order, reaching into the deep-ultraviolet spectral range (Fig. ~\ref{fig:gener}a). The resonant metasurface geometry simultaneously enhanced the local electromagnetic field and increased nonlinear conversion efficiency, enabling compact frequency up-conversion far beyond the capabilities of unpatterned LiNbO$_3$. Importantly, this work highlights the synergy between engineered photonic resonances and intrinsically strong nonlinear materials, establishing ferroelectric oxides as a powerful class of platforms for broadband and efficient harmonic generation at the nanoscale.

A recent advance in nonlinear metaphotonics introduced resonance-gradient metasurfaces as a strategy to overcome the inherently narrowband response of resonant HHG platforms. Jangid et al. demonstrated germanium-based dielectric metasurfaces composed of subwavelength “bone-like” nanoresonators with spatially varying geometry, leading to a continuous spatial gradient of resonance frequencies across the sample~\cite{jangid2024spectral}. This architecture enables spectrally tunable enhancement of HHG by allowing different regions of the metasurface to selectively resonate with different excitation wavelengths (Fig. ~\ref{fig:gener}b). Experimentally, efficient generation of the third and fifth harmonics was observed over a broad tuning range using a mid-infrared laser source, effectively decoupling harmonic enhancement from a single fixed resonance condition.

Another emerging direction is the integration of resonant metasurfaces with van der Waals materials to exploit hybrid photonic–polaritonic interactions for HHG. Sakib et al. demonstrated a silicon-on-sapphire metasurface coupled to a thin layer of hexagonal boron nitride (hBN) \cite{sakib2026polariton}, where a mid-infrared phonon-polariton resonance was co-engineered with a near-infrared photonic resonance to simultaneously enhance the driving field and the emitted fifth harmonic (Fig. ~\ref{fig:gener}c). Numerical calculations predicted an enhancement of the fifth-harmonic yield by more than three orders of magnitude compared with either the metasurface or hBN alone, highlighting the importance of dual-resonance matching between the excitation and harmonic wavelengths. This work establishes hybrid metasurface–van der Waals heterostructures as a promising platform for HHG, demonstrating how coupling photonic and polaritonic modes can unlock new pathways for engineering nonlinear light–matter interactions beyond the capabilities of conventional dielectric metasurfaces.

Graphene is another promising material with broadband tunable nonlinearity in mid-infrared/terahertz spectral region to realise compact ultraviolet sources. High-order harmonics were realised by exploiting strong near-field plasmonic enhancement near doped graphene nanostructures using a terahertz laser source \cite{cox2017plasmon} with microscopic theory proposed to explore the roles of electronic band structure and Coulomb interactions \cite{de2020strong}. Other materials of atomic thickness, such as semiconductor monolayers and transition metal dichalcogenides (TMDs), combining strong nonlinearity with a gapped, valley-selective, spin-orbit-coupled electronic structure, were shown to be suitable for HHG \cite{liu2017high, liu2020polarization, wang2022optical, yue2022signatures}, with dramatic enhancement provided by many-body Coulomb interactions \cite{hader2023coulomb}.

Several classes of exotic materials with strong nonlinear optical responses have not  yet been explored in metaphotonics. Efficient and rich HHG was reported in topological and Weyl materials, attractive for polarization- and helicity-controlled HHG, including non-integer multiples of driving frequency for a topological insulator \cite{schmid2021tunable} and strong odd and even HHG in Weyl semimetal \cite{lv2021high}. Mid-infrared laser excitation can strongly drive the Dirac topological surface states, with bulk and surface contributions to harmonic emission, tunable by the thickness of thin films \cite{li2026disentangling}. Another class includes Mott insulators and strongly correlated materials, exhibiting exotic many-body correlations between charge and spin affecting HHG spectra and unconventional power scaling laws \cite{murakami2018high, uchida2022high}.

\section{Unconventional power dependencies}

The nonlinear optical response of a material is conventionally described within the perturbative regime, where the intensity of the $n$th-order harmonic scales as $I_n \propto I_{\mathrm{pump}}^n$. This simple power-law relationship has long served as a defining signature of nonlinear optical processes, enabling straightforward identification of the harmonic order and providing insight into the underlying nonlinear mechanisms~\cite{Bloembergen1996nolinear,boyd2020nonlinear}. It is worth noticing, however, that low-order harmonics can also have non-perturbative nature and unconventional scaling upon ultrashort laser excitation \cite{hussain2022demonstration}. Resonant metaphotonic platforms made this regime more apparent
~\cite{tonkaev2024even,tonkaev2025unconventionalmet,tonkaev2025unconventional}. The strong local-field enhancement provided by high-$Q$ resonant metasurfaces enables efficient HHG at incident intensities several orders of magnitude lower than those required in conventional bulk materials~\cite{liu2018enhanced,tonkaev2025unconventional}. Under these conditions, the intra-cavity field becomes intrinsically intensity dependent owing to nonlinear modifications of the refractive index, resonance frequency, linewidth, and mode coupling. Consequently, the harmonic yield may exhibit effective power-law exponents that deviate significantly from the nominal harmonic order.

\begin{figure*}[!htbp]
    \begin{center}
    \includegraphics[width=0.9\linewidth]{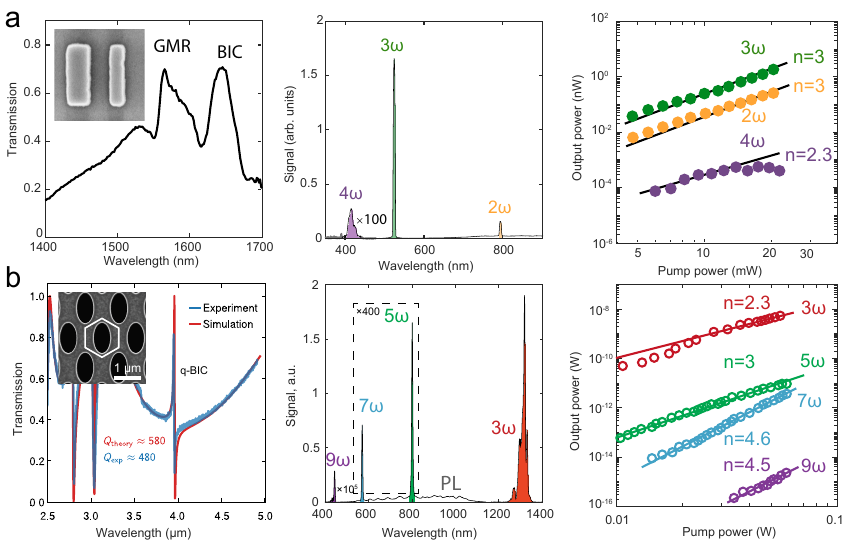}
    \caption{ \textbf{Unconventional power dependencies of high-harmonic generation.} (a) Silicon metasurface supporting quasi-bound states in the continuum and guided-mode resonances in the near-infrared for even- and odd-order high-harmonic generation. Left: Linear transmission spectrum of the metasurface showing the guided-mode resonance and quasi-bound states in the continuum resonance in the near-infrared. The inset shows an SEM image of the metasurface unit cell. Middle: High-harmonic spectra measured under resonant excitation. Right: Power dependencies of the generated harmonics under resonant excitation. Figure adapted with permission from~\cite{tonkaev2024even} \copyright 2024 American Physical Society. (b) Free-standing silicon membrane metasurface supporting a quasi-bound state in the continuum in the mid-infrared for high-harmonic generation. Left: Experimental and simulated linear transmission spectra of the membrane metasurface showing the quasi-bound states in the continuum resonance in the mid-infrared. The inset shows an SEM image of the metasurface unit cell. Middle: High-harmonic spectra measured under resonant excitation. Right: Power dependencies of the generated harmonics under resonant excitation. Figure adapted with permission from~\cite{tonkaev2025unconventional} \copyright 2025 Springer Nature.}
    \label{fig:unconventional}
    \end{center}
\end{figure*}

Beyond enhancing the efficiency of harmonic generation, resonant metasurfaces can fundamentally modify the nonlinear response of materials by altering the effective selection rules and power-dependent behaviour expected from bulk systems. For example, Tonkaev et al. demonstrated the generation of even-order harmonics from centrosymmetric silicon metasurfaces by exploiting engineered resonant electromagnetic modes, including guided-mode resonances and quasi-bound states in the continuum (quasi-BICs), which enable otherwise symmetry-forbidden nonlinear processes under normal incidence~\cite{tonkaev2024even}. The left panel of Fig. ~\ref{fig:unconventional}a presents the linear transmission spectra of the resonant silicon metasurface, revealing two pronounced resonances associated with the guided-mode resonance and quasi-BIC excitation. The corresponding SEM image of the individual meta-atom is shown in the inset. These resonant modes provide strong near-field confinement and substantially enhance the nonlinear interaction volume, resulting in a second-harmonic generation (SHG) response more than three orders of magnitude higher than that of an unstructured silicon film. This enhancement enabled the observation of higher-order even harmonics up to the fourth harmonic, which remain undetectable in nonresonant silicon structures. The middle panel of Fig. ~\ref{fig:unconventional}a shows the measured harmonic spectra under resonant excitation conditions. Interestingly, the enhanced SHG response exhibits a cubic dependence on the incident pump intensity rather than the conventional quadratic scaling expected for a second-order nonlinear process, as shown in the right panel of Fig. ~\ref{fig:unconventional}a. When the excitation wavelength is tuned away from the resonance, the SHG signal recovers the expected quadratic dependence, confirming that the modified scaling originates from the resonantly enhanced optical field rather than from intrinsic material nonlinearities. In contrast, the third-harmonic generation (THG) signal maintains a cubic dependence under both resonant and off-resonant excitation, consistent with conventional third-order perturbative behaviour. Remarkably, the fourth-harmonic generation (FHG) signal exhibits an approximately quadratic dependence under resonant excitation, while it becomes undetectable in the off-resonant regime. These unconventional power dependencies arise from the strong interplay between the nonlinear material response and the intensity-dependent resonant electromagnetic environment, demonstrating that high-$Q$ metasurfaces provide an additional degree of freedom for controlling harmonic generation beyond the intrinsic nonlinear susceptibility of the constituent material.

Going beyond symmetry-enabled harmonic generation, recent studies have revealed that resonant metasurfaces can also fundamentally modify the intensity scaling of high-order harmonic generation. Ref.~\cite{tonkaev2025unconventionalmet} demonstrated that highly resonant free-standing silicon membrane metasurfaces supporting quasi-bound states in the continuum (quasi-BICs) exhibit unconventional non-integer power dependencies in HHG, deviating from the integer-order scaling laws expected from conventional nonlinear optics. The metasurfaces were designed to support a high-Q quasi-BIC resonance in the mid-infrared spectral range. The left panel of Fig. ~\ref{fig:unconventional}b presents the simulated and measured transmission spectra, demonstrating excellent agreement between experiment and theory and confirming the high fabrication quality of the free-standing silicon metasurfaces. The inset shows the SEM image of the metasurface unit cell. The high-$Q$ resonance provides strong local-field enhancement, enabling the observation of harmonic orders up to the ninth order from the silicon membrane. The middle panel of Fig. ~\ref{fig:unconventional}b shows the measured HHG spectra under resonant excitation. In particular, the seventh harmonic exhibits an enhancement of approximately three orders of magnitude compared with an unpatterned free-standing silicon membrane, while the ninth harmonic is observed only in the vicinity of the resonance. Under resonant excitation, the generated harmonic signals exhibit unconventional power-law exponents that are substantially reduced compared with the corresponding harmonic orders predicted by conventional nonlinear optics. The right panel of Fig. ~\ref{fig:unconventional}b shows the measured power dependencies of the harmonic emission under resonant excitation. The third-, fifth-, seventh-, and ninth-order harmonics follow effective scaling exponents of approximately 2.4, 3.0, 4.6, and 4.5, respectively, demonstrating increasingly pronounced deviations from the conventional dependence with increasing harmonic order. In contrast, the third-, fifth-, and seventh-order harmonics generated from the unpatterned free-standing silicon membrane strictly follow the conventional integer-order scaling laws. Numerical modelling revealed that these deviations originate from nonlinear modifications of the resonant state, where Kerr-induced refractive-index changes give rise to self-phase modulation, resonance detuning, and the dynamic redistribution of the intra-cavity electromagnetic field. These findings establish high-$Q$ metasurfaces not only as efficient frequency converters but also as nonlinear photonic platforms capable of fundamentally reshaping the scaling laws governing high-order harmonic generation.

\section{Conclusion and Outlook}

The study of HHG has evolved from phenomena primarily investigated in gases and bulk solids into a rapidly developing area of metaphotonics, where resonant nanostructures provide unprecedented opportunities for engineering strong-field light-matter interactions. As discussed above, the use of subwavelength dielectric resonators and structured metasurfaces has fundamentally transformed the field by enabling efficient HHG at substantially reduced excitation intensities while simultaneously providing new degrees of freedom for controlling the nonlinear response. High-Q resonances, such as bound states in the continuum, have enabled orders-of-magnitude enhancement of the harmonic generation, and they established resonant dielectric nanostructures as an attractive platform for compact extreme nonlinear optics.

Despite remarkable progress demonstrated over the past decade, several important challenges remain. Although resonant nanostructures significantly improve harmonic generation efficiencies, the overall conversion efficiency, particularly for higher-order harmonics, remains below that achieved in conventional bulk systems. Future advances will require resonators supporting stronger field confinement while maintaining high damage thresholds and broad operational bandwidths. In this context, the optimal conditions, when the electronic excitations exhibit pronounced excursions but do not yet result in irreversible structural changes and heating can be avoided by temporal design of laser pulses, remain poorly explored. At the same time, a deeper theoretical understanding based on microscopic description and efficient numerical methods compatible with metasurface dimensions and propagation scales are urgently needed to describe the complex interplay between resonant electromagnetic modes, ultrafast carrier dynamics, and strong-field nonlinearities. Recent observations of unconventional nonlinear power scaling and resonance-induced modifications of harmonic generation suggest that many aspects of strong-field interactions in high-Q nanophotonic systems remain to be understood. So far, the quest for the optimal laser source (wavelength, polarization, orbital angular momentum, etc.), material, and metasurface geometry for selectively generating and manipulating the harmonics of interest below the damage threshold remains open. Furthermore, the signature plateau-and-cutoff HHG spectra have not yet been observed from nanoresonators or metasurfaces. Extending the high-order harmonics to this transition regime by means of metaphotonics might have advantages in resonant control of the shape of the plateau and position of cutoff over HHG in bulk solids, opening new opportunities in attosecond pulse generation and waveform engineering. Ultrafast dynamic control over the electronic trajectory that defines the HHG cutoff by tuning local field amplitude, waveform, polarisation and phase is particularly promising, in view of recent works that demonstrate ultrafast self-action effects on third-order harmonic \cite{sinev2021observation}, ultrafast symmetry breaking to excite quasi-BIC modes \cite{crotti2025g} and switching of nonlinear chirality \cite{lai2025nonlinear} in metasurfaces. Fundamentally, metasurfaces with spatially varying polarization state can reveal local Berry curvature and anisotropic-band effects hidden or averaged out in bulk measurements but also engineer the electronic Hamiltonian itself, particularly in two-dimensional and topological materials. Nonlinear circular dichroism spectroscopy gives additional information about material inversion symmetry properties and how band structure evolves dynamically \cite{chen2020probing,   heinrich2021chiral, bae2022revealing, uzan2024observation}. Strong-field symmetry breaking effects, so far rarely observed in unstructured solids, can become accessible solely by tuning ultrafast light using specially designed metasurfaces.

Looking forward, metaphotonics is poised to evolve beyond simply enhancing existing nonlinear optical phenomena toward creating entirely new regimes of extreme nonlinear optics that are inaccessible in conventional materials. We anticipate some experiments, if demonstrated, would generate a significant and lasting impact in the research community of both strong field physics and metamaterials. 1: Arbitrary and complete, on-demand control of amplitude and phase of the high-order harmonics. Metamaterials will no longer act as a simple field enhancer but rather an engineered toolset that tailors each and individual harmonics and their spectra. 2: Extended set of functionalities for HHG metaphotonics. Drastic improvements in control leads to significant boost of engineering applications, including generation of chiral light, reconfigurable XUV phased-array HHG, etc. 3: Generation, control, and characterization of attosecond pulses. Advances in engineering might bring the full set of the above-mentioned physical processes to be implemented on a chip. Such a demonstration would not only leapfrog frontier developments in attosecond science but its engineered solutions might be used in adjacent light-source technologies such as those used in lithography machines.

Scientific-wise, by simultaneously controlling spatial, temporal, polarization, and symmetry properties of the electromagnetic fields at the nanoscale, resonant nanophotonic structures offer the possibility of engineering the real-space and $k$-space electron trajectory and microscopic mechanisms underlying HHG itself. Continued advances in nanofabrication, ultrafast laser technology, nonlinear materials, and computational design are therefore expected to transform metaphotonics into a versatile platform for integrated attosecond photonics, compact coherent short-wavelength light sources, ultrafast spectroscopy, and on-chip extreme nonlinear optical devices. More broadly, the rapid convergence of nanophotonics and strong-field physics is establishing metaphotonics as a new paradigm for manipulating light–matter interactions at the highest optical intensities, with implications extending well beyond HHG alone.

\begin{acknowledgments}
Y.K. acknowledges a support from the University of Hong Kong via the Hung Hing Ying Distinguished Visiting Professorship in Science and Technology.
He also thanks Maxim Shcherbakov from the University of California Irvine, as well as Zhipei Sun, Nianze Shang, and Zixuan Ning from the Aalto University Finland, for many useful comments, suggestions, and additional references. T.T.L. gratefully acknowledges funding support by the University Grants Committee (17315722); National Natural Science Foundation of China (NSFC)/Research Grants Council of Hong Kong (RGC) Joint Research Scheme (N\_HKU7104/22); and the State Key Laboratory of Optical Quantum Materials.

\end{acknowledgments}

\bibliography{aipsamp}

\end{document}